\documentstyle[12pt]{article}
\newcommand{\beq}{\begin{equation}}
\newcommand{\eq}{\end{equation}}
\newcommand{\beqa}{\begin{eqnarray}}
\newcommand{\eqa}{\end{eqnarray}}
\newcommand{\vect}[1]{\mbox{${\bf #1}$}}
\newcommand{\grvect}[1]{\mbox{\boldmath$#1$}}

\title{Microscopic multicluster description of neutron-halo nuclei
with a stochastic variational method}
\author{K. Varga \\ Physics Department, Niigata University, Niigata 950--21,
Japan \\
and \\
Institute of Nuclear Research of the Hungarian Academy of Sciences, \\
Debrecen, P. O. Box 51, H--4001, Hungary\thanks{Permanent address.}
\and
Y. Suzuki \\ Physics Department, Niigata University, Niigata 950--21, Japan
\and
R. G. Lovas \\
Institute of Nuclear Research of the Hungarian Academy of Sciences, \\
Debrecen, P. O. Box 51, H--4001, Hungary}
\date{Received  June 1993}
\begin{document}
\maketitle
\begin{abstract}
To test a multicluster approach for halo nuclei, we give a unified
description for the ground states of $^6$He and $^8$He in a model
comprising an $\alpha$ cluster and single-neutron clusters.
The intercluster wave function is taken a superposition of terms belonging
to different arrangements, each defined by a set of Jacobi coordinates.
Each term is then a superposition of products of gaussian functions of the
individual Jacobi coordinates with different widths, projected to angular
momenta $l=0$ or 1. To avoid excessively large dimensions and
``overcompleteness", stochastic methods were tested for selecting the
gaussians spanning the basis. For $^6$He, we were able to calculate
ground-state energies that are virtully exact within the subspace defined
by the arrangements and $l$ values, and we found that preselected random sets
of bases (with or without simulated annealing) yield excellent numerical
convergence to this ``exact" value with thoroughly truncated bases. For $^8$He
good energy convergence was achieved in a state space comprising three
arrangements with all $l=0$, and there are indications showing that the
contributions of other subspaces are likely to be small. The $^6$He and $^8$He
energies are reproduced by the same effective force very well, and the
matter radii obtained are similar to those of other sophisticated calculations.
\end{abstract}

\section{Introduction}
The discovery \cite{discovery} of thick neutron clouds or ``halos" in the
surface of some light nuclei has led to a revival of interest in these nuclei,
as is shown by numerous experimental and theoretical studies (see e.g.
\cite{Niigata}). All candidates for halo structure are in the p-shell region,
where systematics can be studied with the shell model, but
the shell model runs into difficulties just in producing halos.
To bring about a halo with a simple-minded enlargement of an
harmonic-oscillator (h.o.) basis would
in most cases require prohibitively large dimensions. If, alternatively,
the single-particle orbits are chosen so as to exhibit the
loosely-bound character of the halo \cite{Bertsch}, then the treatment of the
centre-of-mass (c.m.) motion, of the Pauli principle and
of continuum orbits pose extra problems.

The cluster-orbital shell
model (COSM) \cite{SuzukiIkedaWang} has solved these problems successfully,
and, at least qualitatively, reproduced some known properties of the halo
nuclei $^6$He \cite{SuzukiIkedaWang,Suzuki}, $^8$He \cite{SuzukiIkedaWang} and
$^{11}$Li \cite{TosakaSuzuki}. But the COSM is not necessarily the ultimate
model either. Even for the simplest case of $^6$He, it is difficult to achieve
convergence \cite{Suzuki}. With the number of valence particles increased,
this problem gets aggravated. On the other hand, the nuclei that have two halo
nucleons are often ``borromean": removal of one nucleon prompts a breakup of
the rest by emission of another nucleon. It was shown that
the three-body dynamics characteristic of such a system can be better
described by inclusion of a core+n+n type three-cluster component
\cite{TosakaSuzukiIkeda}.

The alternative models that focus just on the three-body aspects
are the three-body models (see e.g. \cite{Kukulin1,Danilin}).
None of the three-body calculations has, however, achieved unquestionable
success.
%For example, the interactions that describe the $^9$Li+n system
%correctly tend to produce an unbound $^{11}$Li \cite{Kamimura}.
One problem is that the core--nucleon interaction, at least its off-shell
behaviour, is not well-known.
The accuracy of the treatment of antisymmetrization through Pauli projection
(let alone through inclusion of a repulsive interaction \cite{Lehman}) depends
crucially on the off-shell behaviour of the potential. Even if this treatment
is satisfactory for the solution of the dynamical problem, there are extra
three-cluster Pauli effects in the
spectroscopic amplitudes characterizing the disintegration \cite{VargaLovas}.
Moreover, it would be illusive to believe that cores like $^9$Li
are perfectly inert.

All these shortcomings call for fully microscopic treatments that preserve
some elements of the few-body approaches as well. Although microscopic cluster
models are general enough and their techniques are highly
sophisticated \cite{Horiuchi,SuzukiHecht,Okabe}, they have
not been used too much to describe systems containing more than two clusters.
It has been demonstrated recently for an $\alpha$+p+n model of
$^6$Li that a microscopic approach, in which the treatment of the relative
motion is comparable with the three-body approaches is feasible
\cite{CsotoLovasPhysRev}, and similar calculations for the
$^6$He--$^6$Li--$^6$Be isospin triplet have been published in another paper
\cite{Csoto}.

In the present paper we shall apply a multicluster approach to the ground
state (g.s.) of the three-cluster system $^6$He=$\alpha$+n+n and of the
five-cluster system $^8$He=$\alpha$+n+n+n+n. The present approach is
similar to the previous one \cite{CsotoLovasPhysRev}, except for some practical
simplifications, which are compelling for $^8$He, and, for consistency,
are adopted for $^6$He as well. The treatment of $^6$He
is a prerequisite of that of $^8$He because some parameters are to be fitted
to $^6$He. The case of $^6$He is used, at the same time, as a testing ground
of the approximations introduced. We have solved the dynamical problems of
these nuclei, hence obtaining the g.s. energies, and, to test the wave
functions, we have calculated the rms point matter radii. Further
applicaltions of our wave functions will be published in a forthcoming paper.

The model is based on a linear variational method.
The wave function is a superposition of cluster-model terms belonging to
different sets of Jacobi coordinates or ``arrangements", and the only
parameters to be varied are the combination coefficients. Unlike before
\cite{CsotoLovasPhysRev}, we now choose purely central effective
nucleon--nucleon interactions. This leads to a less realistic description
of the relative motion, but it is more consistent with a purely
s-state model of the $\alpha$ intrinsic motion \cite{CsotoLovasPhysRev}.
In consistency with the use of a central force, we allow for fewer values of
the intercluster orbital angular momenta. This is expected to be a fair
approximation because the motion of two halo neutrons relative to each other,
as well as of neutron pairs relative to a core, is dominated by s waves. Just
as before, we expand the relative-motion functions in terms of gaussians of
different widths. To counter the explosion of the basis size, we now have to
introduce special truncation schemes for these expansions. We have examined
systematic as well as stochastic methods for constructing the basis. As will
be seen, the stochastic methods have proved more successful.
It is owing to the stochastic variational procedure that this extremely
versatile model is feasible for practical purposes.

The plan of the paper is as follows. In sect. 2 we outline the model used.
In sect. 3 we introduce various basis truncation schemes
and test them for $^6$He. Sect. 4 comprises the results for $^8$He,
and sect. 5 contains a summary and some conclusions.

\section{The multicluster model}
\label{sec:model}

The model we use is a generalization of the variational multiconfiguration
multicluster approach outlined in ref. \cite{CsotoLovasPhysRev}, and
involves an $\alpha$ cluster and $n$ single-neutron clusters.
It employs a trial function, which is a sum over various
cluster arrangements $\mu$, each associated
with a particular set of intercluster Jacobi coordinates
$\{\grvect{\rho}^{\mu}_1,...,\grvect{\rho}^{\mu}_n\}$.
For $\alpha$+n+n, there are two non-trivially different arrangements,
viz. $\mu$=($\alpha$n)n and $\mu'$=$\alpha$(nn), and the corresponding
intercluster Jacobi coordinates are
\beqa
&\grvect{\rho}_1=\vect{r}_{1}-\vect{r}_{\alpha},
\ \ \ \ \ \ \ \ \ \ \ \ \ \ \
&\grvect{\rho}_2=\vect{r}_{2}
-{\textstyle{1\over 5}}(4\vect{r}_{\alpha}+\vect{r}_{1}),
\nonumber\\
&\grvect{\rho}'_1=\vect{r}_{2}-\vect{r}_{1},
\ \ \ \ \ \ \ \ \ \ \ \ \ \ \
&\grvect{\rho}'_2
={\textstyle{1\over 2}}(\vect{r}_{1}+\vect{r}_{2})-\vect{r}_{\alpha},
\label{3Jacobi}
\eqa
respectively, where $\vect{r}_{\alpha}$ is the position vector of the
$\alpha$ c.m. and $\vect{r}_i$ are those of the extra neutrons. As is seen
in fig.~\ref{Jacobi}a, the Jacobi coordinates for ($\alpha$n)n and $\alpha$(nn)
form a Y-shaped and a T-shaped pattern, respectively, and they will
be referred to as such.
For the system $\alpha$+n+n+n there are four arrangements,
[($\alpha$n)n]n, [$\alpha$(nn)]n, ($\alpha$n)(nn) and $\alpha$[(nn)n],
while for $\alpha$+n+n+n+n there are nine: \{[($\alpha$n)n]n\}n,
\{[$\alpha$(nn)]n\}n, [($\alpha$n)(nn)]n, \{$\alpha$[(nn)n]\}n,
[($\alpha$n)n](nn), [$\alpha$(nn)](nn), ($\alpha$n)[(nn)n],
$\alpha$\{[(nn)n]n\} and $\alpha$[(nn)(nn)]. Fig.~\ref{Jacobi}b shows a few of
the five-cluster arrangements. As an illustration, here we give the
Jacobi coordinates belonging to the TY-shaped arrangement \{[$\alpha$(nn)]n\}n:
\beqa
&\grvect{\rho}_1=\vect{r}_{2}-\vect{r}_{1},
\ \ \ \ \ \ \ \ \ \ \ \ \ \ \ \ \ \ \ \ \ \ \ \ \ \ \ \ \ \ \ \ \ \
\hfill\hfill
&\grvect{\rho}_2=
{\textstyle{1\over 2}}(\vect{r}_{1}+\vect{r}_{2})-\vect{r}_{\alpha},
\nonumber\\
&\grvect{\rho}_3=\vect{r}_{3}
-{\textstyle{1\over 6}}(4\vect{r}_{\alpha}+\vect{r}_{1}
+\vect{r}_{2}),
\ \ \ \ \ \ \ \ \ \ \ \ \ \ \ \hfill
&\grvect{\rho}_4=\vect{r}_{4}
-{\textstyle{1\over 7}}(4\vect{r}_{\alpha}+\vect{r}_{1}
+\vect{r}_{2}+\vect{r}_{3}).
\label{5Jacobi}
\eqa

The nucleon spins are coupled to $S$, and the orbital angular momenta $l_i$
($i=1,...,n$), belonging to the Jacobi coordinates $\grvect{\rho}_i$, are
coupled to $L$. A term pertaining to arrangement $\mu$ and angular
momenta $[S,(l_1...l_n)L]JM$ can be written as
\beq
\Psi ^{\mu}_{S,(l_1...l_n)L}
={\cal A}\left \{\left [\Phi _S
{\cal X}^{\mu}_{(l_1...l_n)L}
(\grvect{\rho}^{\mu}_1,...,\grvect{\rho}^{\mu}_n)
\right ]_{JM}\right \},
\label{trialfnterm}
\eq
where ${\cal A}$ is the intercluster antisymmetrizer,
$\Phi _{SM_S}$ is a vector-coupled product of the $\alpha$ intrinsic wave
function $\Phi ^{\alpha}$ and the $n$ neutron spin-isospin functions
$\Phi^i$, and ${\cal X}^{\mu}_{(l_1...l_n)L}$ is a vector-coupled product of
the intercluster relative functions
$\chi^{\mu}_{il_im_i}(\grvect{\rho}^{\mu}_i)$. The sequence of coupling
is chosen so as to follow the pattern of the Jacobi coordinates.
The intrinsic wave function of the g.s. of the $\alpha$ particle,
$\Phi^{\alpha}$, is an h.o. Slater
determinant divided by the 0s h.o. wave function describing the zero-point
vibration of the $\alpha$ c.m. The h.o. size parameter
$\nu=m\omega/2\hbar$ is determined by
minimization of the $\alpha$ intrinsic hamiltonian. (The spatial factor of
such a function is equal to a product of 0s h.o. functions of three intrinsic
Jacobi coordinates.)

Representing each relative motion belonging to any single set of Jacobi
coordinates by a complete set of functions, one can get a complete basis. The
intercluster functions $\chi^{\mu}_{ilm}(\grvect{\rho}^{\mu}_i)$ are, however,
approximated by finite sums,
\begin{equation}
\chi^{\mu}_{ilm}(\grvect{\rho}^{\mu}_i)
=\sum_{k=1}^{N^{\mu}_i}
C^{\mu}_{ikl}\Gamma^{\mu}_{iklm}(\grvect{\rho}^{\mu}_i),
\label{expansion}
\end{equation}
where $\Gamma^{\mu}_{iklm}$ are h.o. eigenfunctions of radial node number
zero:
\beq
\Gamma^{\mu}_{iklm}(\grvect{\rho}^{\mu}_i)=
\left[{2^{2l+7/2}(\nu^{\mu}_{ik})^{l+3/2}\over\sqrt{\pi}(2l+1)!!}\right]^{1/2}
(\rho^{\mu}_i)^l{\rm exp}[-\nu^{\mu}_{ik}(\rho^{\mu}_i)^2]
Y_{lm}(\hat{\grvect{\rho}}^{\mu}_i).
\label{Gamma}
\eq
The size parameters \{$\nu^{\mu}_{ik}$, $k=1,...,N^{\mu}_i$\} should be
chosen such that the functions $\Gamma^{\mu}_{iklm}$ adequately span the state
space in the relative motion defined by the radial Jacobi coordinate
$\rho^{\mu}_i$. This can always be achieved since any well-behaved functions
can be approximated with any prescribed precision by a combination of such
functions. Because of the loosely-bound nature of these nuclei, more than one
partial wave $l_i$ may be needed for each relative motion, with a rather large
$N^{\mu}_i$ ($\sim 10$) for
the radial motion in each partial wave, which entails an enormously large
basis. Each set of Jacobi coordinates implies a particular asymptotic
configuration, which involves few partial waves just when described in terms
of these particular Jacobi coordinates. Therefore, the dimension may be
reduced by combining the different arrangements:
\beq
\Psi_{(S,L)JM}=\sum_{\mu}\sum_{l_1...l_n}\Psi^{\mu}_{[S,(l_1...l_n)L]JM},
\label{trialfn}
\eq
where the lack of summations over $S$ and $L$ anticipates the choice of a
purely central interaction.

The sets of Jacobi coordinates differ in what physical effects they emphasize.
The Y-shaped and T-shaped branches of the diagrams of fig.~\ref{Jacobi} tend
to magnify the correlation of the neutrons with a ``core" and with each other,
respectively. Thus, for $^8$He the arrangement \{[($\alpha$n)n]n\}n
is best suited to describing a shell-model-like configuration. The
arrangements [$\alpha$(nn)](nn) and $\alpha$[(nn)(nn)] prefer two dineutron
clusters, ($\alpha$n)[(nn)n] and \{$\alpha$[(nn)n]\}n favour
the formation of a three-neutron cluster, $\alpha$\{[(nn)n]n\} seems to
accommodate a four-neutron cluster, while \{[$\alpha$(nn)]n\}n,
[($\alpha$n)(nn)]n and [($\alpha$n)n](nn) are moulded to contain one dineutron
and two neutrons moving around a core in a single-particle manner. It is {\it
a priori} obvious that some of these configurations are very similar to each
other, for instance, the last three arrangements only differ in what the core
is. Thus it seems sufficient to include a few of these configurations. Even
so, the basis subsets belonging to different arrangements do
have large overlaps \cite{CsotoLovasPhysRev,Csoto}, especially for small
distances, which not only allows but, to avoid (numerical) overcompleteness,
also necessitates truncations.

Both the angular momenta and the expansion of the radial functions are to be
truncated. From among highly overlapping subspaces, we shall keep
those with pure $l_i=0$ because, for several clusters, the calculation of the
matrix elements with non-zero $l_i$ is very lengthy.
To introduce a truncation of the expansion of the radial function $\cal X$,
we need to substitute (\ref{expansion}) into $\cal X$ involved in
eq.~(\ref{trialfnterm}):
\beq
{\cal X}^{\mu}_{(l_1...l_n)L\Lambda}=\sum_KD^{\mu}_{K,l_1...l_n}
\left[\Gamma^{\mu}_{1k_1l_1}...\Gamma^{\mu}_{nk_nl_n}\right]_{L\Lambda},
\label{trialexpansion}
\eq
where $K\equiv\{k_1...k_n\}$,
$D^{\mu}_{K,l_1...l_n}=C^{\mu}_{1k_1l_1}...C^{\mu}_{nk_nl_n}$,
and the summation
over $K$ may be truncated such that it will no longer be equivalent to
a sequence of summations over $k_1$,...,$k_n$.

The only variational parameters in the model are the coefficients
$D^{\mu}_{K,l_1...l_n}$. Since $\nu^{\mu}_{ik}$ are not varied,
the bases will be somewhat redundant in the sense that
there will exist smaller bases that would produce the same accuracy.
To keep the variational method linear is, however, advantageous because the
solution is reduced to a diagonalization over a non-orthogonal basis,
and the method is founded on a genuine minimum principle. This implies that
any estimate for any bound-state energy gives an upper bound, and the energies
are lowered monotonously as the basis is extended. Moreover,
if the basis is extended so that, in a limiting case, it fully spans a
well-defined subspace of the state space, the energies will converge to
the exact energies in that subspace. In the foregoing applications
whole sets of variational calculations will be performed with stepwise
extensions of the basis to make estimates for such ``exact" energies by
producing numerical convergence.

We chose the central effective interaction introduced by Thompson, LeMere and
Tang \cite{TLT}, which has the form
\beq
V(i,j)=\left[V_R(r_{ij})+{\textstyle{1\over 2}}(1+P^{\sigma}_{ij})V_t(r_{ij})
+{\textstyle{1\over 2}}(1-P^{\sigma}_{ij})V_s(r_{ij})\right]
{\textstyle{1\over 2}}\left[u+(2-u)P^r_{ij}\right],
\label{force}
\eq
where $r_{ij}=\vert\vect{r}_j-\vect{r}_i\vert$, $P^{\sigma}_{ij}$, $P^r_{ij}$
are the spin- and space-exchange operators, and $V(r_{ij})$ are of gaussian
shape, and we neglected the Coulomb force, which would just shift the energy
scale. This interaction reproduces, at
least qualitatively, the most important low-energy nucleon--nucleon phase
shifts, which implies that it does not bind the dineutron. On the other hand,
by tuning the exchange parameter $u$, we can make the $\alpha$+nucleon system
comply with the borromean character of $^6$He. The parameter $u$ can be
adjusted so as to reproduce the p$_{3/2}$ $\alpha$+n phase shift and the
$^6$He binding energy $E$ with respect to the three-body breakup threshold
simultaneously, with reasonable accuracy. Without having a spin--orbit force,
this is the best one can do. Since in $^8$He the p$_{3/2}$ shell can
accommodate all valence neutrons, this prescription is expected to cause
little error. We chose $u$=1.14, which produces $E=-1.016$ MeV, while the
experimental value is $-0.975$ MeV. The $\alpha$ size parameter, rms point
matter radius and energy implied by this force are $\nu=0.303$ fm$^{-2}$,
1.363 fm and $E_{\alpha}=-25.581$ MeV, respectively.

The matrix elements involving the basis functions are expressed as
integral transforms of matrix elements of Slater determinants, of
shifted Gaussians, projected to total angular momentum $JM$. The technique
applied is similar to that used in our earlier work \cite{CsotoLovasPhysRev}.

\section{Description of $^6$He}
\label{sec:6He}
In the description of a bound-state three-cluster system, like $^6$He, the
use of a large enough basis poses no problem. For the g.s. of $^6$He, a
central force renders the quantum numbers to pure $S=L=0$ and $l_1=l_2$.
Appreciable contributions only come from the subspace defined by the T-shaped
arrangement with $(l_1,l_2)=(0,0)$ (``T$_{00}$ subspace") and by the Y$_{00}$
and Y$_{11}$ subspaces defined analogously. Radial bases of the types of
eq.~(\ref{Gamma}) with parameters
\beq
(\nu^{\mu}_{ik_i})^{-1/2}=0.7\times1.4^{k_i}\ {\rm fm}
\ \ \ \ \ \ \ \ (\ k_i=1,...,10;\ i=1,2)
\label{nu}
\eq
yield energies that are virtually exact within the model space defined by the
arrangement(s) and value(s) of $(l_1,l_2)$ adopted. In table~\ref{6He} we give
these energies along with the corresponding rms point matter radii for the
three subspaces T$_{00}$, Y$_{00}$ and Y$_{11}$ and their unions.

The energies obtained in the individual subspaces are very close to each
other, which shows that they are almost equally important, and yet, when they
are combined, the binding gets stronger moderately. This behaviour is typical
of strongly overlapping subspaces \cite{BDL,CsotoLovasPhysRev,Csoto}. Although,
as can be inferred from the table, none of the subspaces is large enough
by itself, the combination \{T$_{00}$,Y$_{11}$\} does look like that.
The fact that the energy produced by state space \{T$_{00}$,Y$_{11}$\} is
hardly pushed down by inclusion of Y$_{00}$ (with the radius unchanged)
indicates that the ``full" wave function generated in the space
\{T$_{00}$,Y$_{00}$,Y$_{11}$\} is in fact almost entirely contained in subspace
\{T$_{00}$,Y$_{11}$\}.

The observed overlap between the subspaces suggests that the basis is largely
redundant. It would be obviously desirable to reduce its dimension $N$ by
pinpointing redundant elements and dropping them. This is, however, not a
trivial task because the overlaps of each particular element with all others
are diverse. In the following we shall test various tactics of omitting
elements and choosing elements so as to produce less redundant bases. We
should emphasize that, even when the dimension is large (e.g., 300), the bulk
of the computing time is spent on the computation of the matrix elements
rather than on the diagonalization. Therefore, unlike in an earlier work on
basis optimization \cite{Kukulin2}, our aim is to reduce the number of matrix
elements to be computed, and to this end, we are willing to make more
diagonalizations.

The basis of each subspace defined by eq.~(\ref{nu}) can be depicted in a
$(b_1,b_2)=((\nu^{\mu}_1)^{-1/2}$, $(\nu^{\mu}_2)^{-1/2})$ sheet as
is shown in fig.~\ref{bases}a. We see that in each subspace for each Jacobi
coordinate each function $\Gamma$ occurs 10 times while the $\nu$ for the
other Jacobi coordinate runs over all values. It is thus natural to ask
whether one could get a satisfactory basis by a straightforward ``thinning".
Restricting ourselves to the
T$_{00}$ space, we discarded half of the basis elements in a chessboard-like
pattern shown by the open circles in fig.~\ref{bases}a
(i.e. we kept \{$\Gamma^{\mu}_{1k_1l_1}$, $k_1=2,4,...,10$\} for $k_2$ odd and
\{$\Gamma^{\mu}_{1k_1l_1}$, $k_1=1,3,...,9 $\} for $k_2$ even). We obtained
$E=-0.312$ MeV, which implies too much loss ($0.083$ MeV) of binding energy.

Alternatively, one may set up the basis with a stochastic method.
First we present calculations with randomly chosen subsets of the basis
elements defined by eq.~(\ref{nu}), as exemplified by fig.~\ref{bases}b.
Fig.~\ref{discreterandom} shows the energy
versus dimension curves, $E(N)$, with three sets of T$_{00}$ bases constructed
by stepwise inclusion of such elements in fully random sequences. We see that
the convergence is not too fast; by a 50\% truncation, we may lose 200 keV of
binding.

Given the above results, it is interesting to see whether we can gain anything
from a continuously random basis. While in the previous attempt the labels,
which are natural numbers, were picked randomly, now the $\nu$ values
themselves, i.e. two {\it real} random numbers were generated independently.
In our procedure the values of both $\nu$ were limited by imposing
\beq
1\mbox{ fm}\leq\nu^{-1/2}\leq10\mbox{ fm}.
\label{nuspan}
\eq
The bases generated in this way have proved excessively redundant. However,
since the pool of basis states is infinitely large, it is reasonable
to introduce some preselection. With this idea in mind, we introduced an
admission procedure for extending the basis. Since any extension of the basis
lowers the energy, the performance of a basis element can be judged by the
energy improvement it causes. This obviously depends on the basis set in
the previous step. At every step we tried 10 fully random basis states by
adding one at one time in the basis constructed in the previous step, and
admitted only the one that produced the lowest energy. A
$((\nu^{\mu}_{1k_1})^{-1/2},(\nu^{\mu}_{2k_2})^{-1/2})$ set
obtained by such sequential random augmentations of the T$_{00}$ basis is
dispalyed in fig.~\ref{bases}c, and the results of three such sets of
calculations are shown in fig.~\ref{continuousrandom}. All three sets converge
rapidly. At dimensions 30 and 50 the energies are  $-0.367$, $-0.380$,
$-0.376$ MeV and $-0.393$, $-0.395$, $-0.394$ MeV, respectively, while the
corresponding radii are 2.403, 2.442, 2.458 fm  and 2.478, 2.467, 2.474
fm, respectively. Accepting an accuracy of, say 20--30 keV, we can say that,
with such a random basis, a dimension of $\sim$30--40 for a single subspace is
adequate. The reduction of the dimension with respect to the discrete case is
obviously due to the selective sampling.

This method is comparable with stochastic basis selections
that avoid full re-di\-ag\-onal\-iza\-tion at every step. Following
ref.~\cite{Kukulin2}, we repeated the above procedure by replacing
the full diagonalizations with those of $2\times2$ matrices formed by the
model g.s. obtained in the previous step and the state to be tested.
As is seen in fig.~\ref{Kukk}, this procedure converges to the wrong value,
while the same bases with full diagonalizations throughout yield a convergence
almost as good as that resulting from the selection based on repeated
full diagonalizations. Thus, whereas the limiting energy obtained by
sequential $2\times2$  diagonalizations is useless, the resulting basis is
quite good. For problems in which the the diagonalization takes a higher
proportion of the computing time, this procedure is recommendable as
a method for basis construction, but one $N\times N$ diagonalization in the
end is unavoidable.

When the three subspaces are combined, the random selection with full
re-di\-ag\-onal\-iza\-tions results in
fig.~\ref{3subspace}. Three of the four curves were calculated by choosing
40, 20 and 20 basis elements from the subspaces consecutively, with three
different sequences of the subspaces. It is seen that, while in the first
subspace the convergence needs almost 40 elements, in the others it suffices
to take substantially fewer. By subtracting the lengths, along which the
curves are practically horizontal, we see that the fastest convergence could
be attained by starting with subspace Y$_{11}$, which yields the deepest
binding in a single-subspace model (cf. table~\ref{6He}). In calculating the
forth curve, the channel of each basis element was also chosen randomly.
The initial section of this convergence curve is substantially steeper than the
others, but, as we see in fig.~\ref{3subspace}, this is not a gain if we
prescribe an accuracy of 20--30 keV. With the random sequence of channels a 20
keV accuracy is produced by dimension $45$ ($E=-0.997$ MeV).

We have seen that a randomization with preselection may reduce the required
dimension appreciably, but, in general, a random selection does not produce
the optimum. It is possible that one can get a basis that is closer to the
optimum by including all those and only those random elements that surpass a
certain {\it preset} utility level. Utility is not an abstract quality; it
should be understood with respect to the basis generated in the previous step.
This level can thus be defined as a minimum energy improvement $\varepsilon$
required of the inclusion of a would-be basis state. When the admittance is
decided by a utility test, the convergence is automatically signalled by
insistent failure to find further elements that pass the test.

In a pure T$_{00}$ model the utility test with $\varepsilon=0.005$
MeV did not improve the convergence, but in fig.~\ref{bases}d we still show
such a basis. This apparently differs from the preselected random basis in
two respects: it does not contain highly overlapping pairs of basis states
and it contains fewer elements with both $\nu^{-1/2}$ small.
In the full model this method looks definitely superior. In three
runs, all with the sequence \{T$_{00}$,Y$_{00}$,Y$_{11}$\}, the energies and
radii have converged to $-0.999$, $-0.997$, $-0.998$ MeV, and 2.429, 2.422
and 2.427 fm, at dimensions 45, 59, 47, respectively. When no basis element
was found in a subspace to pass the utility test out of
10 consecutive attempts, the procedure was continued with the next
subspace. In the last subspace, however, the calculations were only
terminated when no basis element was found to fulfil the quality
criterium after 100 attempts. The resulting energy and radius as a
function of the dimension are shown in fig.~\ref{qualitycontrolled}.
With $\varepsilon$=0.01 MeV the energy deteriorated by a mere $\sim$20 keV
(it converged to $-0.986$, $-0.984$, $-0.969$ MeV at  $N=$48, 39, 46,
respectively), with a slightly larger scatter in the radii.

Although the energy deepens monotonously and the rate of deepening is mostly
found to decrease with the basis increased, there is no theorem on the latter.
For this reason, due caution should be excercised to avoid declaring
convergence prematurely. In view of this, we modified the method as follows.
After ${\cal N}=10$ attempts failing to reach the utility level, we skipped
the test once, i.e. we admitted a basis element that does not pass it.
Such a procedure is called ``simulated annealing" \cite{annealing}.
The random search for new elements was stopped after
10${\cal N}$=100 consecutive failures to satisfy the utility criterium.
With simulated annealing and parameter $\varepsilon=0.01$ MeV
the energy curves converged to $-0.993$, $-0.996$,
$-0.992$ MeV at dimensions $59$, $57$ and $66$, respectively. Thus the
simulated annealing makes the method less economical, but it does make the
procedure that involves utility tests safer.

Since the energy convergence attainable is not precise in a mathematical
sense, one should check the quality of convergence for the wave function as
well. The overall convergence can be judged from the fluctuations of the
radius, which are acceptable for bases about to produce energy convergence
(see figs.~\ref{continuousrandom}b and \ref{qualitycontrolled}b).
The critical domain of the nucleus is, however, the halo region, which
may be extensive without contributing too much to the binding. For example,
$\beta$-decay observables seem to be affected by the precise shape of the
fall-off of the halo beyond 13 fm \cite{Baye}. We tested
the wave function through examining, in the T$_{00}$ model, the two-variable
function obtained by projecting out, from the wave function
$\Psi^{\rm T}_{[0,(00)0]00}$, the $\alpha$ intrinsic motion and the
$(l_1l_2)=(00)$ angular component of the
relative motions:
\beq
g(R_1,R_2)=R_1R_2
\hbox{${\displaystyle\int}\kern -0.3em d$}\hat{\bf R}_1Y_{00}^*(\hat{\bf R}_1)
\hbox{${\displaystyle\int}\kern -0.3em d$}\hat{\bf R}_2Y_{00}^*(\hat{\bf R}_2)
\langle{\cal A}\{\Phi_0\delta({\bf R}_1-{\grvect\rho}_1)
\delta({\bf R}_2-{\grvect\rho}_2)\}\vert\Psi^{\rm T}_{[0,(00)0]00}\rangle.
\label{spamp}
\eq
This is in fact a spectroscopic amplitude. We examined the effect of including,
in the basis, elements that emphasize the surface region. Starting with the
100-dimensional basis reported on in table~\ref{6He}, we included 10 more
elements, resulting in a change of $-0.00024$ MeV and $+0.0027$ fm
in the energy and radius, respectively. Meanwhile, the norm
square of $g(R_1,R_2)$ changed from $s=$1.3570 to 1.3568, and the
change of $g(R_1,R_2)$ is very small accordingly. At the limit $R_1=R_2=15$
fm, where $g(R_1,R_2)$ is 100 times smaller than its peak value, the change
is 20\%, and rapidly decreases with any of $R_i$ smaller. Therefore, an
integral over any of the two variables, like the function that enters into the
description of $\beta$ decay \cite{Baye}, carries a much smaller error. Thus
the error found seems insignificant even for the halo properties. A random
basis constructed with utility tests ($\varepsilon$=0.005 MeV) and simulated
annealing converges at $N=36$, to yield an 0.008 MeV loss of binding and a
$-0.001$ fm decrease of the radius (with respect to the 100-dimensional case)
and $s=1.3596$. The departure in $g(R_1,R_2)$ for large $R$ values is, however,
smaller than in the previous comparison, obviously because the difference
between these two bases is not enhanced artificially in the asymptotic region.
The slightly larger difference in $s$ is due to a difference in $g(R_1,R_2)$
spread over a larger domain of $(R_1,R_2)$. Thus we can claim that the basis
reduction does not hit the asymptotic region more than any other region, and
the method is as reliable in predicting the halo properties as any other
properties.

To sum up, we have demonstrated that, by constructing the basis with
stochastic sampling, the dimension
required for an accuracy of 20--30 keV can be safely reduced by a factor of
5--6. We note that all results with random bases presented in this section
have been corroborated by at least three different samplings. When there are
more clusters, the basis is expected to be even more redundant, and an even
larger reduction can probably be achieved. At the same time, however, there is
more chance to be bogged down in flat areas, thus utility testing should
always be accompanied with simulated annealing.

\section{Description of $^8$He}
\label{sec:8He}
Since in the $\alpha$+n+n+n+n model considered there are four Jacobi
coordinates, it is beyond our means to construct a basis with a prescription
like eq.~(\ref{nu}) even in a single arrangement. (Then the dimension would be
$10^4$.) The experience with $^6$He suggests, however, that, if numerical
convergence can be reached with a random basis, then there is good hope that
the limiting energy and wave function are approximately equal to the exact
energy and wave function of the model problem. Moreover, it is easy to
ascertain that this is so by examining whether different random bases converge
to the same energy.

Fig.~\ref{8Hepath} shows that calculations with (preselected) random bases
limited by eq.~(\ref{nuspan}) converge rather slowly. The cause of this is to
be found in that the domain of the four-dimensional ``$\nu$-space" included
according to eq.~(\ref{nuspan}) is rather large in comparison with the domain
that gives rise to net attraction between the five clusters. By squeezing the
limits in eq.~(\ref{nuspan}), the convergence can only be speeded up for small
dimensions because the outer regions of this volume also contribute to the
binding appreciably through interference in a configuration mixing. Therefore,
we kept eq.~(\ref{nuspan}) as it is.

Fortunately, however, the convergence can be speeded up by utility-tested
basis states more radically than for $^6$He. In the example of
fig.~\ref{8Hepath}, in subspace \{[$\alpha$(nn)]n\}n with all orbital angular
momenta zero, a preselected random basis leads to an apparent energy
convergence at $E=-2.970$ MeV (measured from the $\alpha$+n+n+n+n five-body
breakup threshold) at dimension  $N=300$. Almost the same energy, $E=-2.961$
MeV, can be reached by utility testing of quality parameter
$\varepsilon=0.005$ MeV, with a simulated annealing of ${\cal N}=10$, at
dimension $N=151$. Therefore, we adopt this prescription as the standard
procedure. The energy is accepted as converged when failure to pass the test
is detected for the $100$th time consecutively.
% This criterium can only be used beyond dimension $N\sim 3$
%because, in case there is just one very good basis element among the first
%few, the $\nu$-domain in which there are would-be basis states that can
%improve the energy more than $\varepsilon$ is rather small; later,
%interference will help to increase the contribution of every single basis
%state up to a point of ``saturation".
In multiconfiguration calculations we limited the
dimension for each subspace to 120, which was either enough to attain
convergence in the above sense or was very close to convergence. In the latter
case it was left to the configurations to be included afterwards to make up
for this slight omission as far as this is within their scope.
We checked all convergence in all cases by at least one more independent set
of calculations, and adopted the lower energy as the calculated model energy.
The agreement between such independent calculations has
always been better than 30 keV in energy and 0.005 fm in radius.

A review of the results is given in table~\ref{8He}. In all configurations
considered the neutron spins are coupled to zero pairwise; in setting up these
pairs we followed the pattern of the Jacobi vectors. All orbital angular
momenta are assumed to be zero except those belonging to
\{[($\alpha$n)n]n\}n$'$, which are coupled as
$\{[(l_1,l_2)l_{12},l_3]l_{12,3},l_4\}L=\{[(1,1)0,0]0,0\}0$.
The configurations considered are chosen so that each qualitatively different
arrangement be represented.

If all angular momenta were included in each arrangement, each
arrangement would span a complete basis by itself. Therefore, a comparison
of the single-configuration calculations can be interpreted as testing the
performance of the single-angular-momentum ansatz in each arrangement.
Alternatively, since each configuration may as well be viewed as exhibiting
certain types of intercluster correlation, these calculations may be
interpreted as exploring these
correlations. The difference between the configurations is surprisingly small.
The good energy of the shell-model-like configuration with pure $l=0$ shows
the enormous flexibility of this model; after all, in the pure shell model a
purely $l=0$ configuration would carry an excitation of $4\hbar\omega$! The
fact that the $\alpha$+(4n)-like configuration $\alpha$\{[(nn)n]n\} performs
so moderately shows that, in the ``all $l=0$ limit", the
neutrons are more closely co-ordinated with the $\alpha$ cluster than with
each other. At first sight, it looks surprising that the two
$\alpha$+(2n)+(2n)-like configurations do not work equally well;
$\alpha$[(nn)(nn)] is less satisfactory obviously because it is closer to an
$\alpha$+(4n)-type configuration like $\alpha$\{[(nn)n]n\}. The arrangement
($\alpha$n)[(nn)n] is a representative of $\alpha$+(3n)+n-like formations, and,
not surprisingly, is energetically disfavoured with $l=0$ for similar reasons.
The $\alpha$+(2n)+n+n-like arrangement yields a deep binding, which
suggests that a configuration with two closely co-ordinated and
two loosely co-ordinated neutrons, which is consistent with a two-neutron
halo, has a large weight. The best result is, however, furnished by the
\{[($\alpha$n)n]n\}n$'$ configuration, which is not entirely unexpected since
it contains, as it were, the best single $^6$He configuration, Y$_{11}$.

Some of the multiconfiguration calculations show that the single-configuration
models leave some room for improvements. The component $\alpha$\{[(nn)n]n\}
does not seem to add very much to the best single configuration with $l=0$,
but the combination of the two favoured configurations,
\{[$\alpha$(nn)]n\}n and [$\alpha$(nn)](nn), do improve on the energy
substantially. The third favoured $l=0$ configuration, \{[($\alpha$n)n]n\}n, in
\{[$\alpha$(nn)]n\}n+[$\alpha$(nn)](nn)+\{[($\alpha$n)n]n\}n still has an
appreciable effect. Pilot calculations show, however, that a forth
configuration has no significant effect. In particular, although the
configuration containing $l=1$, \{[($\alpha$n)n]n\}n$'$, is the best by
itself, it scarcely deepens the binding when added to the superposition
\{[$\alpha$(nn)]n\}n+[$\alpha$(nn)](nn)+\{[($\alpha$n)n]n\}n. This statement
lacks preciseness because the $l\not=0$ ingredients make full-fledged
computations of such types prohibitively slow. We managed to include 17
basis states of subspace \{[($\alpha$n)n]n\}n$'$ and that improved the energy
by a mere 0.0015 MeV.

It was shown in ref.~\cite{KLBD} that it is possible to reproduce the energies
of $^6$Li and $^8$Be in the cluster model with the same central force rather
well, but $^8$He is much more complicated than $^8$Be. Now we see that, just
as $^6$He, the nucleus $^8$He is somewhat overbound by the force chosen. By
modifying the mixing parameter from $u=1.14$ to $1.135$, the binding of $^6$He
reduces to -0.940 MeV (with $r=2.462$ fm), while the
\{[$\alpha$(nn)]n\}n+[$\alpha$(nn)](nn)+\{[($\alpha$n)n]n\}n model gives
$-3.041$ MeV (with $r=2.272$ fm) for $^8$He. A linear interpolation shows
that the force that puts the energy of $^6$He right yields $-3.204$ MeV
for $^8$He, which is in excellent agreement with experiment, indeed.
It is also remarkable that, while the separation energy of $^8$He changes
significantly, there is virtually no change in the radius. This suggests
that the energy change, and thus the energy itself, is smeared all over the
single-particle degrees of freedom.

\section{Conclusion}
\label{sec:conclusion}

We have argued that a thorough understanding of neutron-halo nuclei calls
for a microscopic multicluster approach. In this paper we have demonstrated
that such an approach is feasible. For the time being, we have adopted a
central nucleon--nucleon interaction and a pure h.o. configuration for the
$\alpha$ particle, and described the g.s. of the $\alpha$+n+n and the
$\alpha$+n+n+n+n systems. We formulated the problem in a model space
consisting of a few arrangements of the $\alpha$ and single-neutron
clusters moving with orbital angular momenta of 0 or 1 around each other.
This assumption takes shape in well-defined model state spaces, and we
wished to get nearly exact g.s. solutions in these spaces.
Our approach uses a linear variational method, and
the main technical problem was the construction of a basis that
contains this ``exact" solution.

These multicluster problems are few-body problems aggravated by the
complications coming from the internal structure of $\alpha$.
The exact solution of a nuclear five-body problem is beyond the
present-day technical limits, and a most successful model-free approximation
to such problems applies stochastic techniques in the variational calculus
combined with similar (Monte Carlo) integration techniques (see e.g.
\cite{Pandharipande}). The method we have now introduced for multicluster
problems only applies random sampling techniques at the level of the
variational approach, viz. in constructing the trial function. This approach
is still feasible because all matrix elements involved can be calculated
analytically, even though they need substantial amount of computation after
all.

Drawing on former results, for $^6$He we adopted a three-component state space,
and constructed a basis in a systematic way. Invariance of the results
against amendments to the basis has proven that the basis contains the ``exact"
g.s. to a good approximation. Having such an ``exact" solution, we tested
ideas of constructing smaller bases that still encompass the ``exact"
solution. It was found that step-by-step enlargements of the basis by
qualified random elements can be used to construct smaller
but almost as good bases. What qualifies a basis function is its contribution
to the energy when added in the basis constructed in the previous step.
We tried to select candidates for basis states both by comparing them with
each other (i.e. we singled out the best among a few in every step) and with
an absolute scale (i.e. we singled out all that contribute more than a preset
value). Since the latter type of selection (``utility testing") might
terminate itself too soon, we allowed temporary lapses in the rigour of
admittance (``simulated annealing").

For $^8$He the same stochastic methods of constructing compact bases
converged, and different methods or different random sets led to the same
limiting energies. We thus have good reason to believe that these limiting
energies are in fact the ``exact" energies. While for $^6$He any of the
methods works equally well, for $^8$He the utility testing proved definitely
superior. To describe $^8$He, it seems to suffice to include a
three-component state space, which is a superposition of an
$\alpha$+(2n)+n+n-type, an $\alpha$+(2n)+(2n)-type and a shell-model-like
arrangement.

In spite of the simplicity of the interaction used, we managed to reproduce
the g.s. energies of $^6$He and $^8$He simultaneously with an accuracy of 100
keV, which is a remarkable success. Nevertheless, the calculated matter radii
of both nuclei undershoot the most recent experimental data. This result is
consistent with that of Cs\'ot\'o \cite{Csoto} for $^6$He, who used
non-central terms along with the same central force, and extended the state
space accordingly. These radii may be accounted for by the $\alpha$-particle
ingredient being too small; for the $\alpha$ wave function involved in the
multicluster model has the same size parameter as producing 1.363 fm for the
radius of the free $\alpha$ particle, which is $\sim 8$\% smaller than the
empirical value. This and other physical implications of the model will be
discussed in a forthcoming paper.

\vskip 12pt
This work was supported by the OTKA grants No. 3010 and F4348 (Hungary) and
by a Grant-in-Aid for Scientific Research (No. 03640265) of the Ministry of
Education, Science and Culture (Japan). K. V. is grateful to the Physics
Department, Niigata University, for their kind hospitality. Y. S. would like
to thank the Japan Society for the Promotion of Science for a fellowship.
The help of Dr. A. Cs\'ot\'o in testing the computer codes used in this work
is gratefully acknowledged.

\newpage
{\footnotesize
\begin{table}[h]
\caption{Energies and rms radii of $^6$He in different model spaces}
\begin{tabular}[t]{lr@{}lr@{}lr@{}lr@{}lr@{}lr@{}lr@{}lr@{}l}
\hline\\
Subspace &
\multicolumn{2}{c}{T$_{00}$} &
\multicolumn{2}{c}{Y$_{00}$} &
\multicolumn{2}{c}{Y$_{11}$} &
\multicolumn{2}{c}{\{T$_{00}$,Y$_{00}$\}} &
\multicolumn{2}{c}{\{T$_{00}$,Y$_{11}$\}} &
\multicolumn{2}{c}{\{Y$_{00}$,Y$_{11}$\}} &
\multicolumn{2}{c}{\{T$_{00}$,Y$_{00}$,Y$_{11}$\}} &
\multicolumn{2}{c}{Experiment} \\
\\\hline\\
$E$ (MeV) & --0.&395 & --0.&387 & --0.&435 & --0.&944
& --1.&015 & --0.&605 & --1.&016 & --0.&975 \\
$r$ (fm) & 2.&455 & 2.&441 & 2.&451 & 2.&411 & 2.&443 & 2.&469 & 2.&442
&2.&4 $^{\rm a}$)\\
\\\hline\\
\end{tabular}
$^{\rm a}$) 2.48$\pm 0.03$, ref. \cite{Heradiusold}; 2.33$\pm 0.04$, ref.
\cite{Heradiusnew}.
\label{6He}
\end{table}
}

\begin{table}[h]
\caption{Energies and rms radii of $^8$He in different model spaces}
\begin{tabular}[t]{lcr@{}lr@{}l}
\hline\\
Subspace $^{\rm a})$ & $N$ & \multicolumn{2}{c}{$E$ (MeV) $^{\rm b})$}
                     & \multicolumn{2}{c}{$r$ (fm) $^{\rm c})$} \\
\\ \hline\\
\{[($\alpha$n)n]n\}n &       153 & --2.&509 & 2.&341 \\
\{[($\alpha$n)n]n\}n$'$; \{[(1,1)0,0]0,0\}0
                     &       130 & --3.&033 & 2.&322 \\
\{[$\alpha$(nn)]n\}n &       161 & --2.&961 & 2.&305 \\
$[\alpha$(nn)](nn)   &       163 & --2.&676 & 2.&278 \\
$\alpha$[(nn)(nn)]   &       176 & --1.&167 & 2.&188 \\
($\alpha$n)[(nn)n]   &       220 & \multicolumn{2}{c}{unbound} \\
$\alpha$\{[(nn)n]n\} &       157 & --1.&368 & 2.&196 \\
\{[$\alpha$(nn)]n\}n+[$\alpha$(nn)](nn)
                     &       211 & --3.&292 & 2.&306 \\
\{[$\alpha$(nn)]n\}n+$\alpha$\{[(nn)n]n\}
                     &       223 & --3.&027 & 2.&225 \\
\{[$\alpha$(nn)]n\}n+[$\alpha$(nn)](nn)+\{[($\alpha$n)n]n\}n
                     &       303 & --3.&395 & 2.&271 \\
\{[$\alpha$(nn)]n\}n+[$\alpha$(nn)](nn)+$\alpha$\{[(nn)n]n\}
                     &       311 & --3.&321 & 2.&317 \\
\\\hline\\
\end{tabular}
\\
$^{\rm a})$ The neutron spins are coupled to 0 pairwise, and all orbital
angular
momenta are zero except those belonging to \{[($\alpha$n)n]n\}n$'$, which are
coupled as $\{[(l_1,l_2)l_{12},l_3]l_{12,3},l_4\}L=\{[(1,1)0,0]0,0\}0$. \\
$^{\rm b})$ Experiment: --3.112 MeV \\
$^{\rm c})$ Experiment: 2.52$\pm 0.03$, ref. \cite{Heradiusold};
2.49$\pm 0.04$, ref. \cite{Heradiusnew}.
\label{8He}
\end{table}

\newpage
\begin{figure}[h]
\vspace{0cm}
\caption{Schematic diagrams depicting all possible sets of Jacobi coordinates
for $^6$He (a) and some possible sets for $^8$He (b). The $\alpha$
cluster and the neutrons are represented by large and small dots, respectively.
\label{Jacobi}}
\vspace{0cm}
\caption{The basis elements of the subspace T$_{00}$ represented by full or
open circles in the $(b_1,b_2)=((\nu^{\mu}_1)^{-1/2},(\nu^{\mu}_2)^{-1/2})$
sheet. The basis of
$(\nu^{\mu}_{ik_i})^{-1/2}=0.7\times1.4^{k_i}$ fm $(k_i=1,...,10;\ i=1,2)$ with
the open circles omitted in the chessboard-like thinning (a); a random subset
(full circles) of the previous basis (b); a random basis with preselection
(c); a utility-tested basis (d). Note the logarithmic scale!
\label{bases}}
\vspace{0cm}
\caption{Random paths of energy convergence in the T$_{00}$ subspace on
the pool of discrete basis elements of
$(\nu^{\mu}_{ik_i})^{-1/2}=0.7\times1.4^{k_i}$ fm $(k_i=1,...,10;\ i=1,2)$.
\label{discreterandom}}
\vspace{0cm}
\caption{Convergence in the T$_{00}$ subspace (horizontal line: exact value)
with (preselected) random bases in energy (a) and in the rms point matter
radius (b). \label{continuousrandom}}
\vspace{0cm}
\caption{Convergence in the T$_{00}$ subspace (horizontal line: exact value)
with consecutive $2\times2$ diagonalizations (short dashes), with full
diagonalizations on the same bases (long dashes) and with a random basis
constructed with full diagonalizations (full line). \label{Kukk}}
\vspace{0cm}
\caption{Energy convergence with (preselected) random bases in the
\{T$_{00}$,Y$_{00}$,Y$_{11}$\} space. The four curves differ in the
sequence of subspaces switched on. Sequence T$_{00}$,Y$_{00}$,Y$_{11}$:
short-dashed curve; sequence Y$_{00}$,Y$_{11}$,T$_{00}$: medium-dashed curve;
sequence Y$_{11}$,T$_{00}$,Y$_{00}$: long-dashed curve;
basis elements randomized between subspaces: full curve;
horizontal line: exact value. \label{3subspace}}
\vspace{0cm}
\caption{Convergence of energy (a) and rms radius (b) produced by three
random sequences of bases constructed with utility tests of parameter
$\varepsilon=0.005$ MeV in the \{T$_{00}$,Y$_{00}$,Y$_{11}$\}
space. Horizontal line: exact value. \label{qualitycontrolled}}
\vspace{0cm}
\caption{Convergence of the $^8$He energy with a (preselected) random basis
(dashed curve) and with a basis constructed with utility tests of
$\varepsilon=0.005$ MeV and ${\cal N}=10$ (full curve)
in the subspace \{[$\alpha$(nn)]n\}n restricted by all orbital angular
momenta set to zero.
\label{8Hepath}}
\end{figure}

\end{document}